\documentclass[12pt, draftclsnofoot, journal, onecolumn]{IEEEtran}
\usepackage{blindtext}
\usepackage[T1]{fontenc}
\usepackage{amsmath}
\usepackage{amssymb}
\usepackage{bm}
\usepackage{mdwmath}
\usepackage{graphicx}
\usepackage{xcolor}
\usepackage[caption=false,font=footnotesize,subrefformat=parens]{subfig}
\usepackage{makecell}
\usepackage{multirow}
\usepackage{acronym}
\usepackage[noadjust]{cite}
\usepackage{url}

\usepackage{algpseudocode}
\newcounter{MYalgorithmic}

\newcounter{MYitem}[MYalgorithmic]

\newcommand{\upperroman}[1]{\uppercase\expandafter{\romannumeral#1}}

\newcommand{\myexp}{\mathrm{e}}
\newcommand{\MYIEEEmembership}[1]{\IEEEmembership{\normalsize #1}}
\begin{document}
\ifCLASSOPTIONonecolumn
	\typeout{The onecolumn mode.}
	\title{\LARGE Ultra-Reliable and Low-Latency Communications for Connected Vehicles: Challenges and Solutions}
	\author{{\normalsize Haojun~Yang,~\MYIEEEmembership{Student Member,~IEEE},
			Kan~Zheng,~\MYIEEEmembership{Senior Member,~IEEE}, 
			Kuan~Zhang,~\MYIEEEmembership{Member,~IEEE},
			Jie~Mei,~\MYIEEEmembership{Student Member,~IEEE},
			and~Yi~Qian,~\MYIEEEmembership{Fellow,~IEEE}}
	\thanks{Manuscript received April 2, 2019; revised XXX.}
	\thanks{Haojun~Yang, Kan~Zheng and Jie~Mei are with the Intelligent Computing and Communication ($ \text{IC}^\text{2} $) Lab, Wireless Signal Processing and Network (WSPN) Lab, Key Laboratory of Universal Wireless Communication, Ministry of Education, Beijing University of Posts and Telecommunications (BUPT), Beijing, 100876, China (E-mail: \textsf{yanghaojun.yhj@bupt.edu.cn; zkan@bupt.edu.cn; meijie.wspn@bupt.edu.cn}).}
	\thanks{Haojun Yang, Kuan~Zhang and Yi~Qian are with the Department of Electrical and Computer Engineering, University of Nebraska-Lincoln, Omaha, NE 68182, USA  (E-mail: \textsf{haojun.yang@unl.edu; kuan.zhang@unl.edu; yi.qian@@unl.edu}).}}
\else
	\typeout{The twocolumn mode.}
	\title{Ultra-Reliable and Low-Latency Communications for Connected Vehicles: Challenges and Solutions}
	\author{Haojun~Yang,~\IEEEmembership{Student Member,~IEEE},
			Kan~Zheng,~\IEEEmembership{Senior Member,~IEEE}, 
			Kuan~Zhang,~\IEEEmembership{Member,~IEEE},
			Jie~Mei,~\IEEEmembership{Student Member,~IEEE},
			and~Yi~Qian,~\MYIEEEmembership{Fellow,~IEEE}
	\thanks{Manuscript received April 2, 2019; revised XXX.}
	\thanks{Haojun~Yang, Kan~Zheng and Jie~Mei are with the Intelligent Computing and Communication ($ \text{IC}^\text{2} $) Lab, Wireless Signal Processing and Network (WSPN) Lab, Key Laboratory of Universal Wireless Communication, Ministry of Education, Beijing University of Posts and Telecommunications (BUPT), Beijing, 100876, China (E-mail: \textsf{yanghaojun.yhj@bupt.edu.cn; zkan@bupt.edu.cn; meijie.wspn@bupt.edu.cn}).}
	\thanks{Haojun Yang, Kuan~Zhang and Yi~Qian are with the Department of Electrical and Computer Engineering, University of Nebraska-Lincoln, Omaha, NE 68182, USA  (E-mail: \textsf{haojun.yang@unl.edu; kuan.zhang@unl.edu; yi.qian@@unl.edu}).}}
\fi

\ifCLASSOPTIONonecolumn
	\typeout{The onecolumn mode.}
\else
	\typeout{The twocolumn mode.}
	\markboth{IEEE Network}{Yang \MakeLowercase{\textit{et al.}}: Title}
\fi

\maketitle

\ifCLASSOPTIONonecolumn
	\typeout{The onecolumn mode.}
	\vspace*{-50pt}
\else
	\typeout{The twocolumn mode.}
\fi
\begin{abstract}
With the rapid development of communications and computing, the concept of connected vehicles emerges to improve driving safety, traffic efficiency and infotainment experience. Due to the limited capabilities of sensors and information processing on a single vehicle, vehicular networks (VNETs) play a vital role for the realization of connected vehicles. To achieve the closer cooperation on the road, ultra-reliable and low-latency communications (URLLCs) become the indispensable components of future VNETs. However, traditional wireless networks are designed with the objective of maximizing spectral and energy efficiency regardless of adequate consideration of URLLCs. To this end, this article investigates the challenges and solutions of conceiving URLLC VNETs dedicated for connected vehicles. Specifically, the underlying application scenarios are first presented with various vehicular communication patterns. Then, the URLLC requirements and potential challenges are discussed for VNETs. We also discuss the solutions to establish URLLC VNETs from the aspects of physical and media access control layers, including the optimization frameworks for latency and reliability, as well as the case studies of employing such frameworks. Finally, several open research directions are pointed out.
\end{abstract}

\ifCLASSOPTIONonecolumn
	\typeout{The onecolumn mode.}
	\vspace*{-10pt}
\else
	\typeout{The twocolumn mode.}
\fi

\IEEEpeerreviewmaketitle

\ifCLASSOPTIONonecolumn
	\ifCLASSOPTIONjournal
		\typeout{The onecolumn journal mode.}
		\newpage
	\fi
\fi


\section{Introduction}
\label{sec:Introduction}

\IEEEPARstart{W}{ith} the remarkable advancements in communications and computing, the concept of connected vehicles is becoming a reality. Connected vehicles can reduce traffic accidents caused by the drivers' incorrect or careless operations, and improve travel efficiency by adopting efficient time management and reasonable route planning~\cite{Siegel2018}. Furthermore, \textbf{autonomous} vehicles, as the future ultimate form of connected vehicles, are more in need of robust communication networks exchanging real-time information to offer the assistance of control~\cite{adv2015}. As a result, vehicular networks (VNETs) promote two categories of vehicular applications for accelerating the realization of connected vehicles, i.e., safety-related and non-safety-related applications. Among diverse performance indicators of vehicular applications, latency and reliability play a key role to plan the blueprint for connected vehicles.

Due to the highly dynamic change of VNET topology, satisfactory services may not be provided only through a single wireless access network. In general, long term evolution (LTE)-related systems can offer a wider communication coverage area for all vehicles, while dedicated short range communication (DSRC) system can support near-real-time safety messages distribution in local areas. Hence, heterogeneous VNETs (HetVNETs) integrate the advantages of DSRC and LTE, and significantly improve the experience of driving and infotainment. As the requirements of vehicular communications become higher, DSRC system is gradually being abandoned because of the poor deployment of roadside infrastructures. By contrast, thanks to the rapid development of LTE-based vehicle-to-everything (V2X) communications, LTE-related systems are regarded as the most promising solutions for future VNETs~\cite{3gpp885}. Nevertheless, traditional cellular networks are designed with the objective of maximizing spectral and energy efficiency while ignoring the latency and reliability requirements. Therefore, to efficiently support connected vehicles on the road, the fifth generation (5G) concept proposes a novel usage scenario named ultra-reliable and low-latency communications (URLLCs)~\cite{itu2015}.

However, latency and reliability are difficult to be uniformly modeled and optimized within one theoretical framework, because of the hierarchical and complex architecture of the network. Establishing the VNETs that meet rigorous URLLC requirements is still challenging for connected vehicles. For example, both overtaking in freeway and cooperative collision avoidance in urban intersection benefit from the perceptual and control information received from roadside infrastructure or other vehicles via V2X communications. They generally require an end-to-end latency within a few milliseconds and a reliability in terms of error probability down to $ 10^{-5} $. Moreover, eliminating the effects of actual traffic environments to the URLLC performance is another challenge for future VNETs. For instance, the vehicle mobility makes the fast fading effects of radio channels more serious and the network topology more complex. Guaranteeing the timely and effective radio resource allocation for URLLCs is also difficult in the hostile environment. Without the assistance of URLLC VNETs, users may refrain from accepting connected vehicles, which would remain as a far-off futuristic idea. To this end, it is paramount to develop the optimization frameworks, and conceive various URLLC techniques for future VNETs, assisting connected vehicles.

In this article, we investigate the challenges and solutions of URLLC VNETs dedicated for connected vehicles. Specifically, we first put forward the underlying application scenarios with four V2X communication patterns. Then, the communication requirements and potential challenges of establishing URLLC VNETs are discussed in detail. In order to address these challenges, we present the solutions for conceiving URLLC VNETs from the aspects of physical (PHY) and media access control (MAC) layers, including the optimization frameworks for latency and reliability, as well as the case studies of vehicular communication systems under freeway and urban scenarios based on such frameworks. Finally, we point out several future research directions.

The remainder of this article is organized as follows. Several underlying application scenarios of V2X are first presented in Section~\ref{sec:Scenarios}. Then, section~\ref{sec:Challenges} discusses the potential challenges for establishing URLLC VNETs. The solutions and future research directions are subsequently put forward in Section~\ref{sec:Solutions} and~\ref{sec:Directions}, respectively. Finally, conclusions are stated in Section~\ref{sec:Conclusions}.

\section{Underlying Application Scenarios}
\label{sec:Scenarios}

\ifCLASSOPTIONtwocolumn
\begin{figure}[!t]
	\centering
	\includegraphics[scale=0.3]{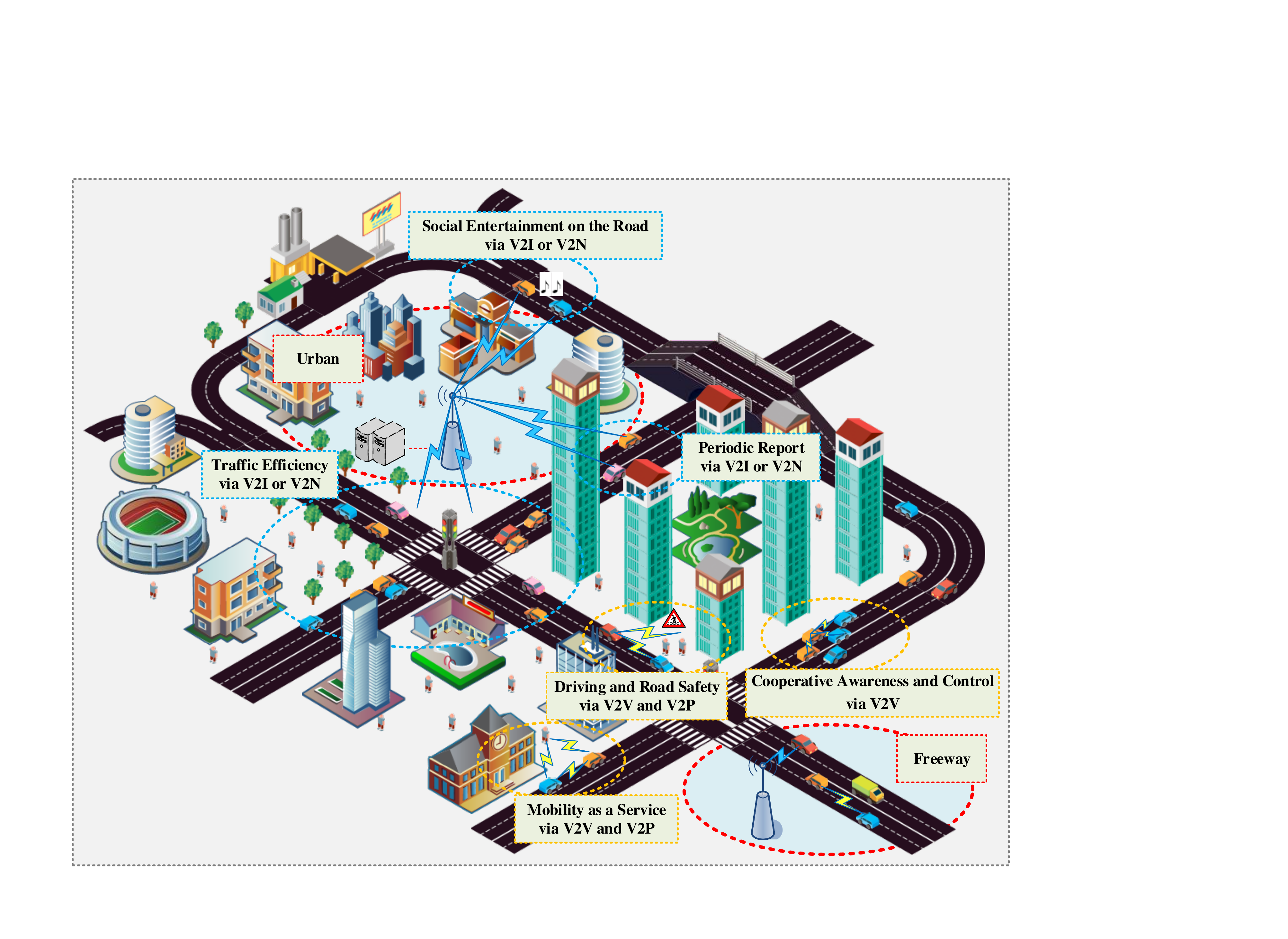}
	\caption{Illustration of underlying application scenarios.}
	\label{fig_scenario}
\end{figure}
\fi

\ifCLASSOPTIONonecolumn
	\begin{figure*}[!t]
		\centering
		\includegraphics[scale=0.45]{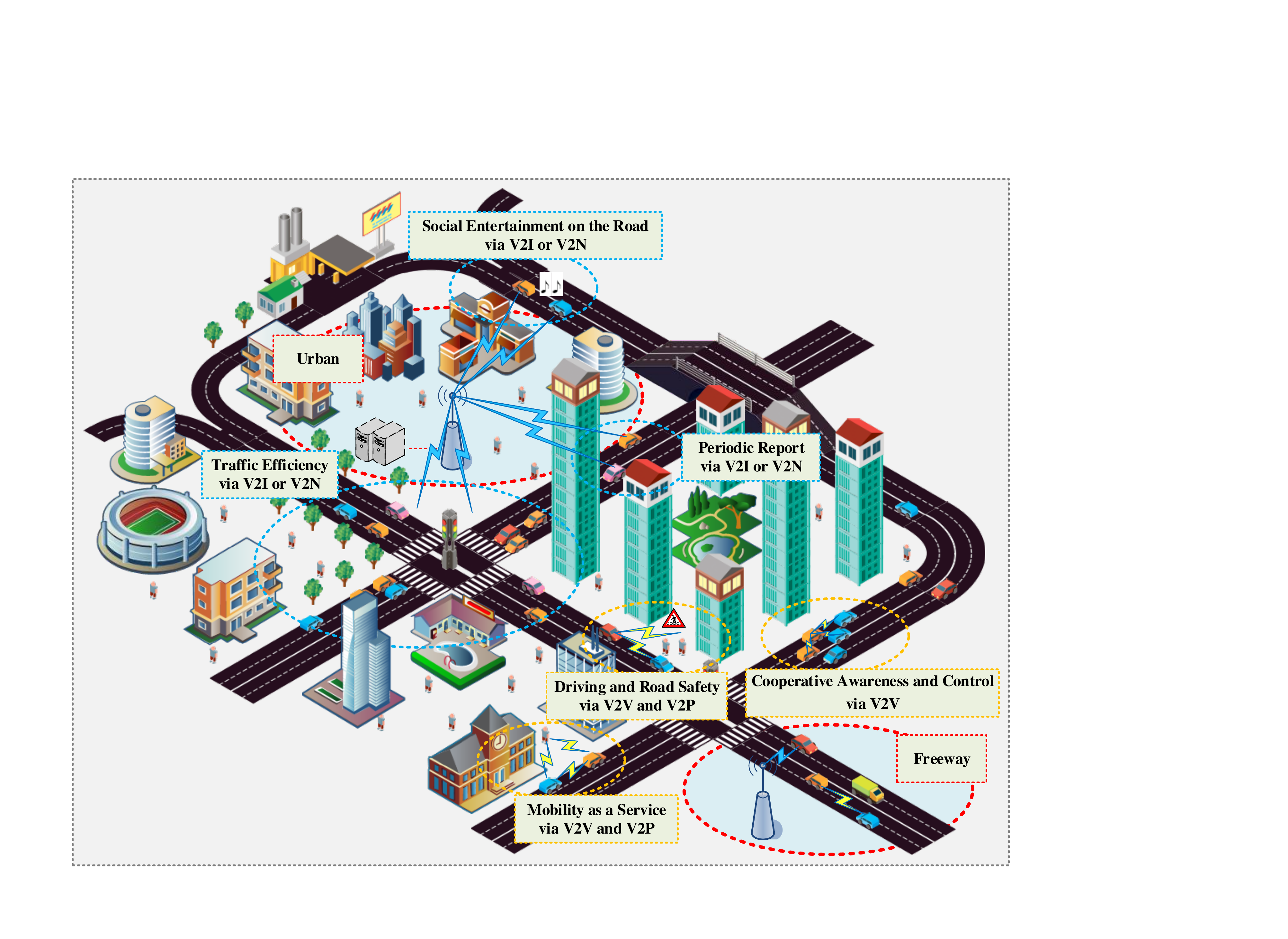}
		\caption{Illustration of underlying application scenarios.}
		\label{fig_scenario}
	\end{figure*}
\fi

\ifCLASSOPTIONtwocolumn
	\begin{table*}[!t]
		\renewcommand{\arraystretch}{1.3}
		\setlength{\extrarowheight}{1pt}
		\centering
		\caption{Summary of communication requirements.}
		\label{tab_requirement}	
		\begin{tabular}{ c | l | c | c | c }
			\hline
			\textbf{Pattern} & \textbf{Typical scenario} & \textbf{Latency} & \textbf{Reliability} & \textbf{Data rate} \\
			\hline
			\hline
			\multirow{3}{*}{V2V and V2P} & Driving and road safety & Ultra-low ($ \leqslant $ 1 ms) & Ultra-high ($ \leqslant 10^{-5} $) & Low \\
			\cline{2-5}
			& Cooperative awareness and control & Ultra-low ($ \leqslant $ 1 ms) & Ultra-high ($ \leqslant 10^{-5} $) & Medium \\
			\cline{2-5}
			& Mobility as a service & Medium & Medium & High \\
			\hline
			\multirow{3}{*}{V2I and V2N} & Traffic efficiency & Low ($ \geqslant $ 1 ms \& $ \leqslant $ 5 ms) & High ($ \geqslant 10^{-5} $ \& $ \leqslant 10^{-3} $) & Low \\
			\cline{2-5}
			& Periodic report & Low ($ \geqslant $ 1 ms \& $ \leqslant $ 5 ms) & High ($ \geqslant 10^{-5} $ \& $ \leqslant 10^{-3} $) & Medium \\
			\cline{2-5}
			& Social entertainment on the road & Medium & Medium & Ultra-high \\
			\hline
		\end{tabular}
	\end{table*}
\fi

In the VNETs, V2X communications are classified into four different patterns, i.e., vehicle-to-vehicle (V2V), vehicle-to-infrastructure (V2I), vehicle-to-network (V2N), and vehicle-to-pedestrian (V2P). As shown in Fig.~\ref{fig_scenario}, the underlying application scenarios are divided into two categories (orange and blue) according to operational areas, one of which is based on V2V and V2P patterns, and the other is based on V2I and V2N patterns.

\subsection{V2V and V2P Patterns-Based Scenarios}
The characteristics of V2P patterns are similar to those of V2V, thus they will be discussed together in this section. V2V and V2P patterns are executed only within local areas and at fine time resolution in order to ensure driving safety and improve driving experience. Three main types of scenarios are studied as follows.

\subsubsection{Driving and Road Safety}
By taking advantage of URLLC VNETs, traffic accidents can be significantly reduced. For instance, in order to avoid driving blind spots, vehicles can broadcast overtaking messages to surrounding vehicles via cooperative V2V communications. As traditional vehicular applications, some of the other use cases are involved in driving and road safety but not limited to these, such as lane changing (via V2V), collision avoidance (via V2V/V2P), etc. Obviously, these use cases require a rigorously low latency and a high reliability.

\subsubsection{Cooperative Awareness and Control}
The capabilities of sensors on a single vehicle is limited. Therefore, with the aid of multi-vehicle cooperative awareness, the reliability of connected vehicles can be effectively improved. Furthermore, we can integrate different control functionalities of vehicles into a global control plane, which is convenient to monitor and manage them. Finally, V2P pattern in this scenario is the most reasonable and efficient way for pedestrians and vehicles to interact with each other.

\subsubsection{Mobility as a Service}
In contrast to the above-mentioned use cases, mobility as a service (MaaS) is a non-safety-related use case. In general, MaaS offers users a solution to find the most appropriate means of transportations and telecommunications. For example, in order to enjoy more efficient and comfort travel, pedestrians can broadcast actual requirements to surrounding vehicles via V2P communications, and then the desirable vehicles can carry them quickly. Moreover, vehicles can find pilots to make an accurate navigation through V2V/V2P communications. Finally, owe to the characteristics of flexible mobility, vehicles can be used as mobile relays to meet the communication requirements in the poor coverage areas.

\subsection{V2I and V2N Patterns-Based Scenarios}
Different from V2V and V2P, V2I and V2N patterns mainly focus on large-scale cooperation, which aims to enhance traffic efficiency and  improve infotainment experience. Three main types of scenarios are discussed here.

\subsubsection{Traffic Efficiency}
Traffic congestion and environmental pollution have become important global issues due to the low traffic efficiency. Generally, the macroscopic model describes the average behavior of some vehicles at specific locations and instances, treating the road-traffic similarly to fluid dynamics. Hence, congestion and pollution can be mitigated by the efficient control of macroscopic flow via V2I/V2N communications. Furthermore, V2I/V2N communications can deal with broadcast storm caused by emergency vehicles, and improve the efficiency of emergency services.

\subsubsection{Periodic Report}
The periodic report is divided into manual and autonomous driving. For manual driving vehicles, both machinery state messages and road condition information should be reported to infrastructures, which helps department of transportation comprehensively grasp the traffic situation. With respect to autonomous vehicles, besides the messages mentioned-above, a huge volume of data generated by sensors should be sent additionally. Then the services of predictive maintenance can be offered by powerful background servers and advanced analysis algorithms.

\subsubsection{Social Entertainment on the Road}
This scenario can be regarded as enhanced mobile broadband (eMBB) services shifted to future VNETs for connected vehicles~\cite{Luan2015}. In spite of this, some particular features in vehicular environments should be considered, such as regional characteristics and self-organization characteristics.

\section{Potential Challenges for VNETs}
\label{sec:Challenges}

\subsection{Communication Requirements}

\ifCLASSOPTIONonecolumn
\begin{table*}[!t]
	\renewcommand{\arraystretch}{1.3}
	\setlength{\extrarowheight}{1pt}
	\centering
	\caption{Summary of communication requirements.}
	\label{tab_requirement}	
	\begin{tabular}{ c | l | c | c | c }
		\hline
		\textbf{Pattern} & \textbf{Typical scenario} & \textbf{Latency} & \textbf{Reliability} & \textbf{Data rate} \\
		\hline
		\hline
		\multirow{3}{*}{V2V and V2P} & Driving and road safety & Ultra-low & Ultra-high & Low \\
		\cline{2-5}
		& Cooperative awareness and control & Ultra-low & Ultra-high & Medium \\
		\cline{2-5}
		& Mobility as a service & Medium & Medium & High \\
		\hline
		\multirow{3}{*}{V2I and V2N} & Traffic efficiency & Low & High & Low \\
		\cline{2-5}
		& Periodic report & Low & High & Medium \\
		\cline{2-5}
		& Social entertainment on the road & Medium & Medium & Ultra-high \\
		\hline
	\end{tabular}
\end{table*}
\fi

In order to achieve the ultimate goal of connected vehicle on the road, 3GPP develops V2X communications from three stages. Both Stage 1 and 2 utilize LTE-related techniques to enhance the support for VNETs. While Stage 3 plans to develop V2X based on 5G new radio (NR). For URLLCs, 3GPP also declares that user plane latency should be 0.5 ms both for uplink and downlink. On the other hand, a general URLLC reliability requirement for one transmission of a packet is $ 10^{-5} $ for 32 bytes with a user plane latency of 1 ms~\cite{3gpp913}. In the VNETs, different scenarios require various latency and reliability. For example, the safety-related applications have a much higher priority to the non-safety-related ones. Hence, they require very rigorous latency and reliability. The non-safety-related applications are not sensitive to latency and reliability, but require a high data rate. Various requirements of latency and reliability are summarized in Table~\ref{tab_requirement}.

\subsection{Potential Challenges}

\subsubsection{Lack of Optimization Frameworks for Different Scenarios}

URLLCs are the indispensable components of future VNETs dedicated for connected vehicles. As the LTE-based V2X communications develop rapidly, LTE-related systems are seen as the most promising solutions for future VNETs. However, traditional cellular networks are designed to maximize spectral and energy efficiency without adequate consideration of URLLCs. Furthermore, latency and reliability are difficult to be uniformly modeled and optimized within one theoretical framework, due to the hierarchical and complex architecture of the network. In order to optimize the URLLC performance of VNETs more specifically, we must consider latency and reliability from different aspects. Generally, the encountered types of latency may be classified according to the hierarchical architecture of the network. PHY layer mainly focuses on reducing the transmission (air interface) latency, while MAC layer pays more attention to the queueing (scheduling) latency. Therefore, it is necessary but challenging to first study the optimization frameworks for latency and reliability from the aspects of PHY and MAC layers.

\subsubsection{Influence of Mobility Models}

Based on the optimization frameworks studied, we can further optimize the performance of URLLC VNETs. A key factor affecting the URLLC performance is the road-traffic model. According to the traffic theory, the road-traffic model is divided into macroscopic and microscopic. In detail, the macroscopic model describes the average behavior of a certain number of vehicles at specific locations and instances, treating the road-traffic similarly to fluid dynamics. By contrast, the microscopic model focuses on the specific behavior of each individual entity (such as vehicle or pedestrian), it is more sophisticated than the macroscopic model. Based on the characteristics of the road-traffic model, the macroscopic model is generally used for the performance evolution of PHY layer, while the microscopic model is suitable for MAC layer. Hence, modeling the effects of traffic on communications accurately is a challenge for studying URLLC techniques.

\subsubsection{Influence of VNET Deployment}

The actual traffic environment is also an important perspective. Typically, VNETs are deployed in the freeway and urban as illustrated in Fig.~\ref{fig_scenario}. In the freeway scenario, vehicles can travel at the desired speed (free flow) or the synchronized speed (synchronized flow)~\cite{adv2015,3gpp885}. While in the urban scenario, the behavior of vehicles and pedestrians is regulated by traffic lights, thus the driving speed is much lower than that of the freeway. Due to the relative high speed of VNETs, the fast fading effects of radio channels are quite serious, which significantly deteriorate the quality of V2X communications. Moreover, the high vehicle density of the urban makes network topology more complex compared to the freeway. Therefore, ensuring URLLCs under these scenarios is a critical challenge for future VNETs.

\section{Solutions to Conceive URLLC VNETs}
\label{sec:Solutions}

To tackle the challenges mentioned above, the optimization frameworks for latency and reliability are studied in this section from the aspects of PHY and MAC layers. Based on the frameworks discussed, three case studies are illustrated for conceiving URLLC VNETs.

\subsection{Frameworks for PHY Layer}

In the scenarios, such as \textit{periodic report}, both the ergodic and outage capacity are reasonable performance metrics, because the packet size is typically large. However, the assumption of large packet size can not be met in the short-packet safety-related scenarios, like \textit{driving and road safety}. Hence, we study the solutions from the perspective of large-packet transmission and short-packet transmission.

\subsubsection{Large-Packet Transmission}

In order to efficiently optimize the performance of URLLC VNETs, we need to first reveal the fundamental tradeoff between latency and reliability.

\paragraph{Fundamental Tradeoff}
The Shannon-Hartley theorem quantifies the error-free channel capacity at which information can be transmitted over a band-limited channel in the presence of noise. However, in wireless communications, what we generally discuss is the outage capacity due to channel fading, rather than the channel capacity (ergodic capacity). Based on orthogonal frequency division multiplexing (OFDM)-related systems, let $ T_\text{s} $ and $ f_\text{s} $ denote the time duration and frequency bandwidth of a resource element (RE). Then the tradeoff between transmission latency $ L $ and reliability $ P_\text{out} $ is given by
\begin{align}
\label{outage}
P_\text{out}&= \Pr \left\{ \left\lfloor \dfrac{L}{T_\text{s}} \right\rfloor \left\lfloor \dfrac{B}{f_\text{s}} \right\rfloor \log_2\left( 1+\rho \right)<C \right\} \notag \\
&=\int_{ - \infty }^{\rho _\text{th}} \rho f(\rho) \text{d}\rho,
\end{align}
where $ B $ denotes the total bandwidth, $ C $ represents the packet size (the number of bits) and $ \rho_\text{th} =2^q-1, q = C / \left( \lfloor \frac{L}{T_\text{s}} \rfloor \lfloor \frac{B}{f_\text{s}} \rfloor \right) $~\cite{Long2018}. According to different operational cases, $ \rho $ can be modeled as the signal-to-noise ratio (SNR) in the interference-free case. By contrast, $ \rho $ represents the signal-to-interference-plus-noise ratio (SINR). $ f(\rho) $ is the probability density function (PDF) of $ \rho $, which is related to channel characteristics.

\paragraph{Case Study~\upperroman{1}---Performance of Outage Probability and Latency with Multi-Antenna Diversity}

\begin{figure}[!t]
	\centering
	\subfloat[Non-correlation case.]{\includegraphics[scale=0.3]{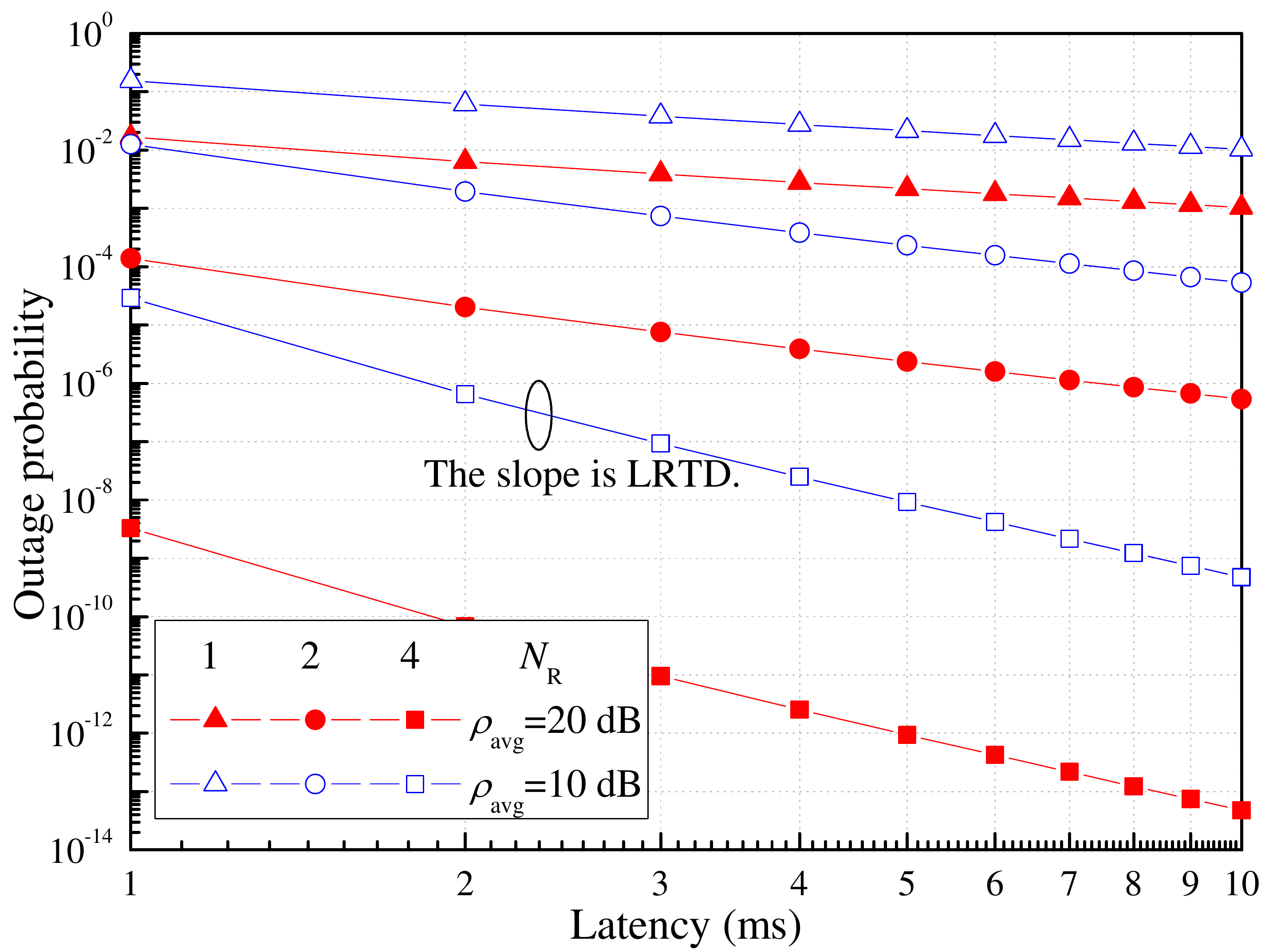}\label{fig_noncorrelation}}
	\hfil
	\subfloat[Correlation case, where the correlation coefficient is 0.5.]{\includegraphics[scale=0.3]{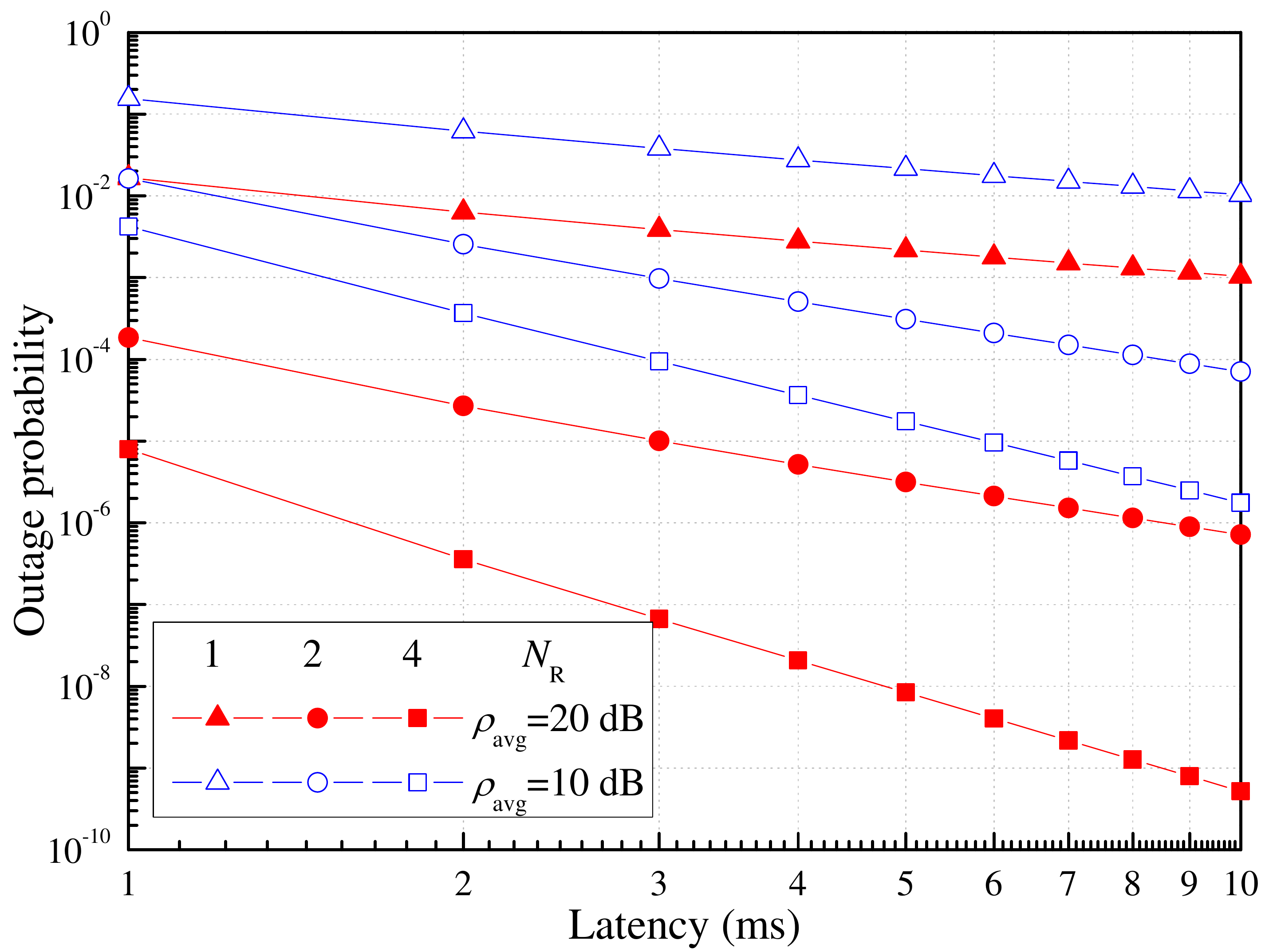}\label{fig_correlation}}
	\caption{Case Study~\upperroman{1}---Performance of outage probability and latency with multi-antenna diversity, where $ C=256 $ bits (32 bytes), $ B=180 $ kHz and $ T_\text{s}f_\text{s}=1 $.}
	\label{fig_case1}
\end{figure}

Multiple-input multiple-output (MIMO) composed of diversity and multiplexing, is the most important technique for OFDM-related systems. The method of ensuring a high reliability is deemed to be the multi-antenna diversity due to the diversity and array gains on the SNR (or SINR). According to Equ.~\eqref{outage}, the effects of multi-antenna to latency and reliability are first studied. All fast fading channels obey independent and identically distributed (i.i.d.) complex Gaussian random distribution with the average SNR $ \rho_\text{avg}=10 $ and $ 20 $ dB. Perfect channel state information (CSI) is also available here. For simplicity, only the case of single-input multiple-output (SIMO) with maximum ratio combining (MRC) is illustrated.

Fig.~\ref{fig_case1} shows the tradeoff between latency and reliability with different numbers of antennas under the log-log plot. As shown in Fig.~\subref*{fig_noncorrelation}, the outage probability monotonically decreases with the increase of latency, which means that there is a tradeoff between latency and reliability. Furthermore, the outage probability decreases with the increasing number of antennas. Given the latency and $ N_\text{R} $, the outage probability decreases as the average SNR increases. The latency has the same trend with the outage probability in the average SNR. Therefore, high average SNR is conducive to enhance latency and reliability. Finally, the correlation of channel is considered. Compared with Fig.~\subref*{fig_noncorrelation}, Fig.~\subref*{fig_correlation} illustrates that the correlation can reduce the reliability. On the other hand, in order to quantify the slope of the logarithm-scaled curve, a new metric named the latency-reliability tradeoff degree (LRTD) can be defined as $ d = -\lim_{L \rightarrow \infty} \log P_\text{out}/ \log L $, which is similar to the system diversity order~\cite{Long2018}.

\subsubsection{Short-Packet Transmission}

Similarly dealing with the large-packet transmission, we first study the fundamental tradeoff between latency and reliability, and then show the specific case study.

\paragraph{Fundamental Tradeoff}
Both the ergodic and outage capacity can be approached at the cost of excessive coding length (latency), i.e, $ C(\rho) = \mathbb{E}\left[ B\log_2(1+\rho) \right] = \lim_{\epsilon \rightarrow 0} C_{\epsilon}(\rho,\epsilon) = \lim_{\epsilon \rightarrow 0}\lim_{n \rightarrow \infty} R(\rho,n,\epsilon) $. For the operational wireless systems, they are reasonable performance metrics, because the packet size is typically large. Nevertheless, the assumption of large packet size can not be satisfied in some short-packet URLLC scenarios, such as \textit{driving and road safety}. Fortunately, during the last few years, significant progress has been made to satisfactorily address the problem of approximating $ R(\rho,n,\epsilon) $~\cite{Polyanskiy2010},~\cite{She2017}, i.e.,
\begin{align}
\label{FBL}
R(\rho,L,\epsilon) = \mathbb{E}\left\lbrace B\left[ \log_2(1+\rho)-\sqrt{\dfrac{V}{LB}}Q^{-1}(\epsilon) \right] \right\rbrace,
\end{align}
where $ Q^{-1}(\cdot) $ denotes the inverse of the Gaussian $ Q $-function and $ V $ is the so-called channel dispersion. For a complex quasi-static fading channel, the channel dispersion is given by $ V=\left( 1- 1/\left( 1+\rho \right)^2 \right) \left( \log_2 \myexp \right)^2 $. $ L $ is the transmission latency, while $ LB $, which is also referred to as the blocklength of channel coding, represents the number of transmitted symbols. When $ LB $ is high enough, the approximation~\eqref{FBL} approaches the ergodic capacity. Compared with the ergodic capacity, the approximation~\eqref{FBL} also implies that the rate reduction is proportional to $ 1/\sqrt{LB} $ when aiming to meet a specific error probability at a given packet size. $ \epsilon $ can be seen as the proxies for reliability. Equ.~\eqref{FBL} clearly shows the tradeoff among latency, reliability and capacity.

\paragraph{Case Study~\upperroman{2}---Performance of Freeway V2I URLLC System~\cite{Yang2019}}

\ifCLASSOPTIONonecolumn
	\typeout{The onecolumn mode.}
	\begin{figure}[!t]
		\centering
		\subfloat[Freeway V2I system based on macroscopic road-traffic model.]{\includegraphics[scale=0.42]{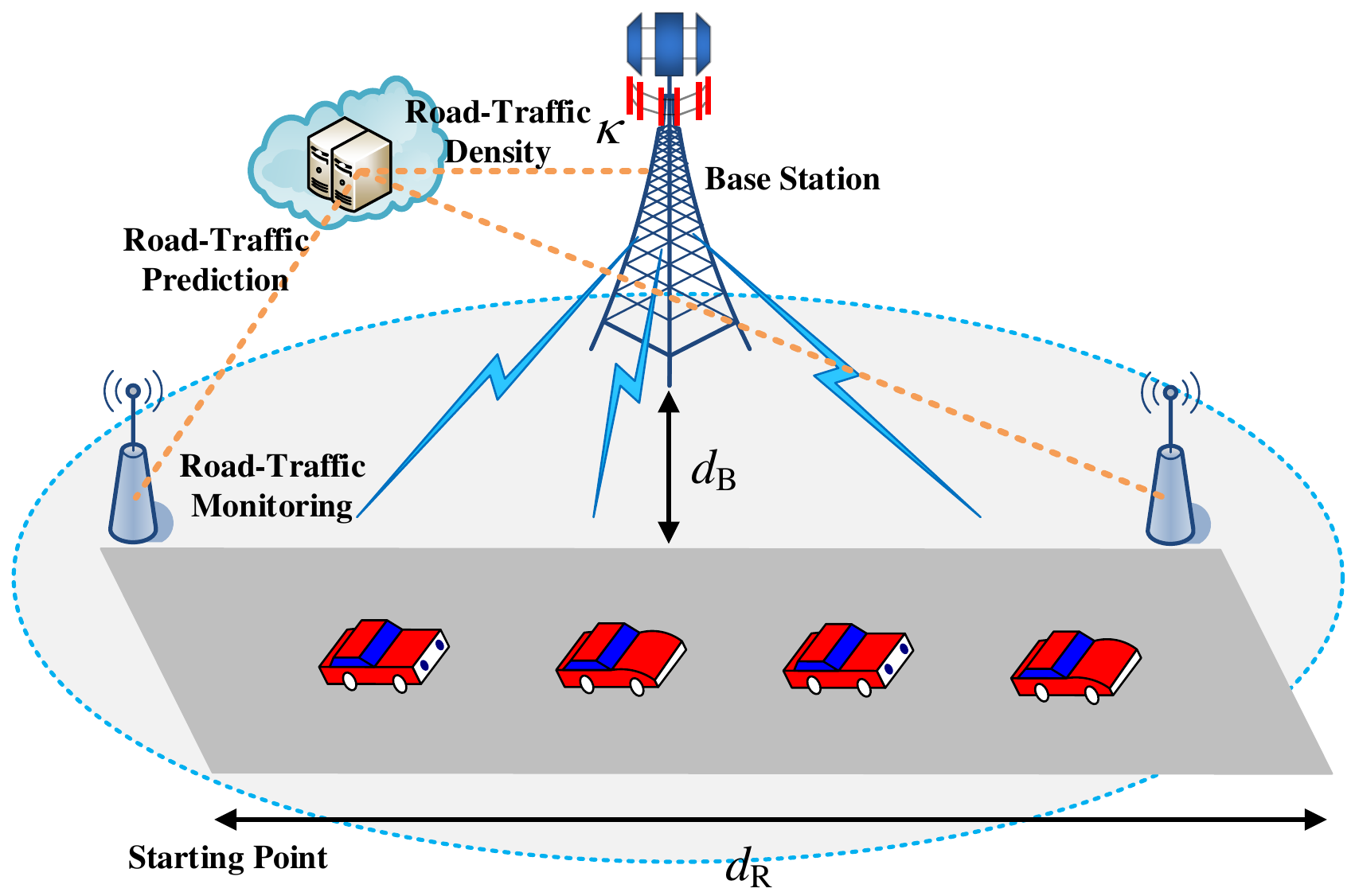}\label{fig_systemcase2}}
		\hfil
		\subfloat[Tradeoff among maximum achievable rate, transmission latency and reliability based on the MF precoder, with $ \kappa = 0.05 $, $ M =300 $, $ B = 200~\text{kHz} $ and the total available power is 10 dBW.]{\includegraphics[scale=0.45]{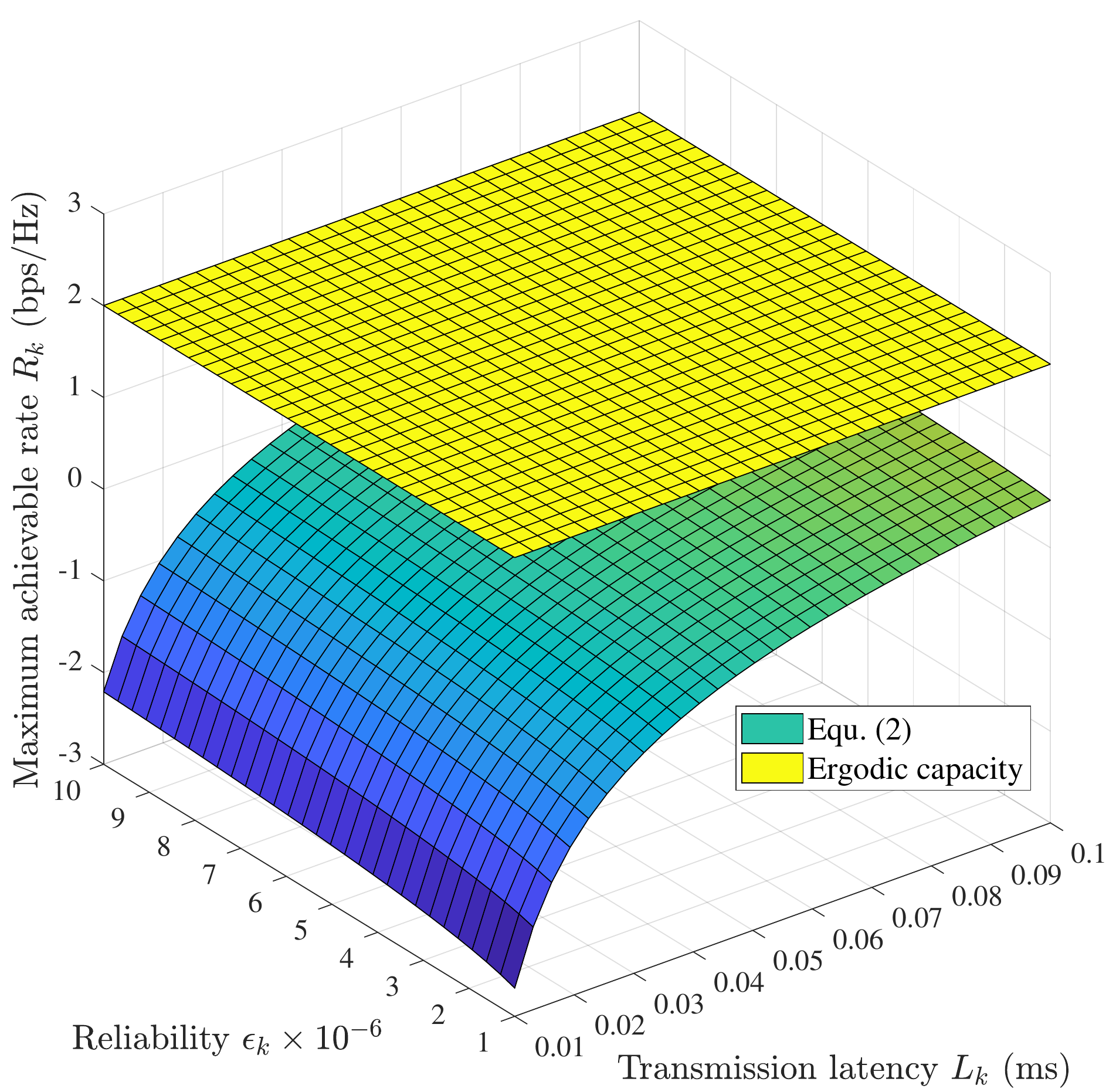}\label{fig_RLR_MF}}
		\hfil
		\subfloat[Maximum transmission latency versus road-traffic density with $ M =300 $, $ R_k(\rho,L,\epsilon) = 100~\text{kbps} $ and $ \epsilon_k = 10^{-6} $ for all vehicular users.]{\includegraphics[scale=0.3]{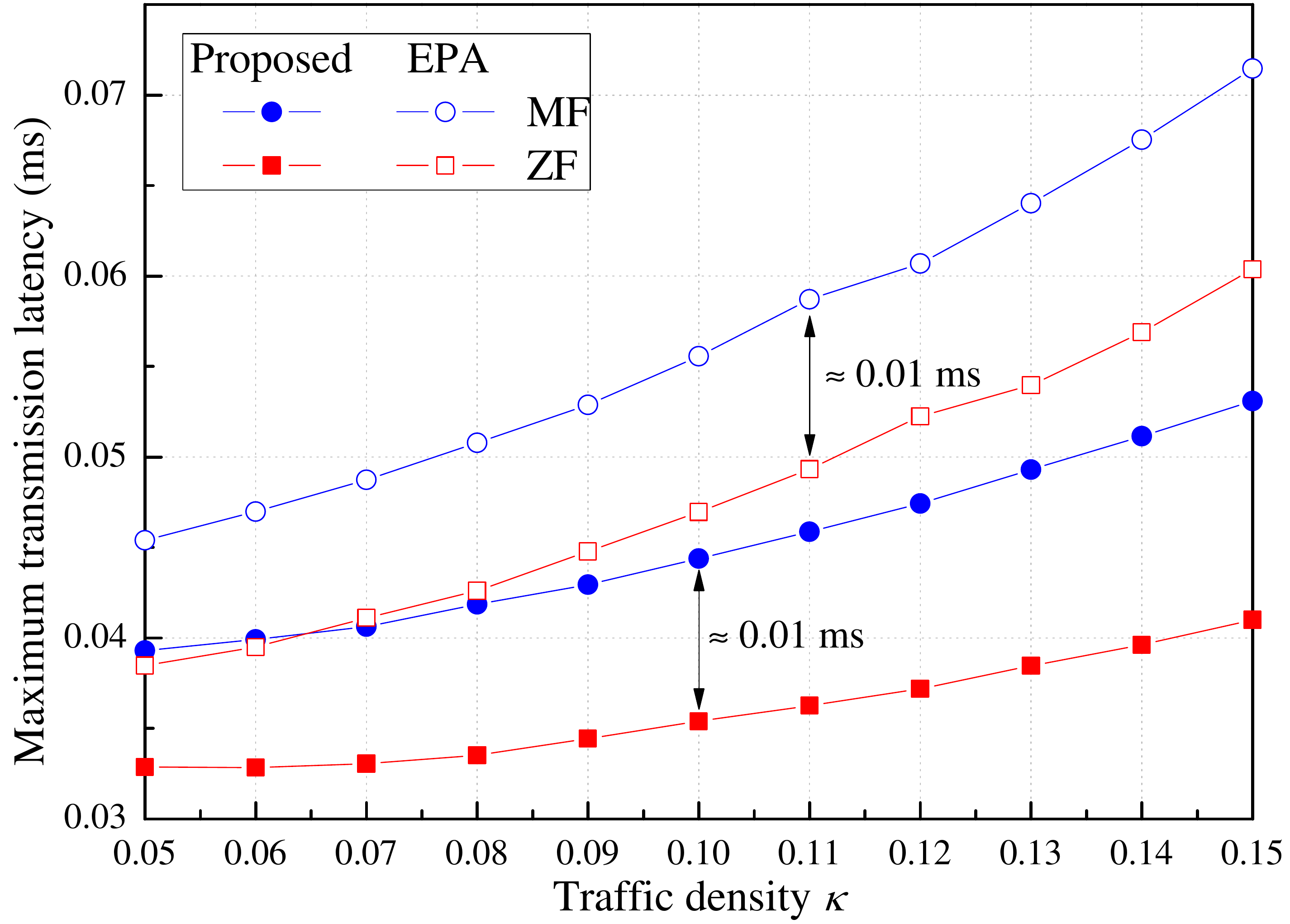}\label{fig_rhovsL}}
		\caption{Case Study~\upperroman{2}---Performance of freeway V2I URLLC system.}
		\label{fig_case2}
	\end{figure}
\else
	\typeout{The twocolumn mode.}
	\begin{figure}[!t]
		\centering
		\subfloat[Freeway V2I system based on macroscopic road-traffic model.]{\includegraphics[scale=0.4]{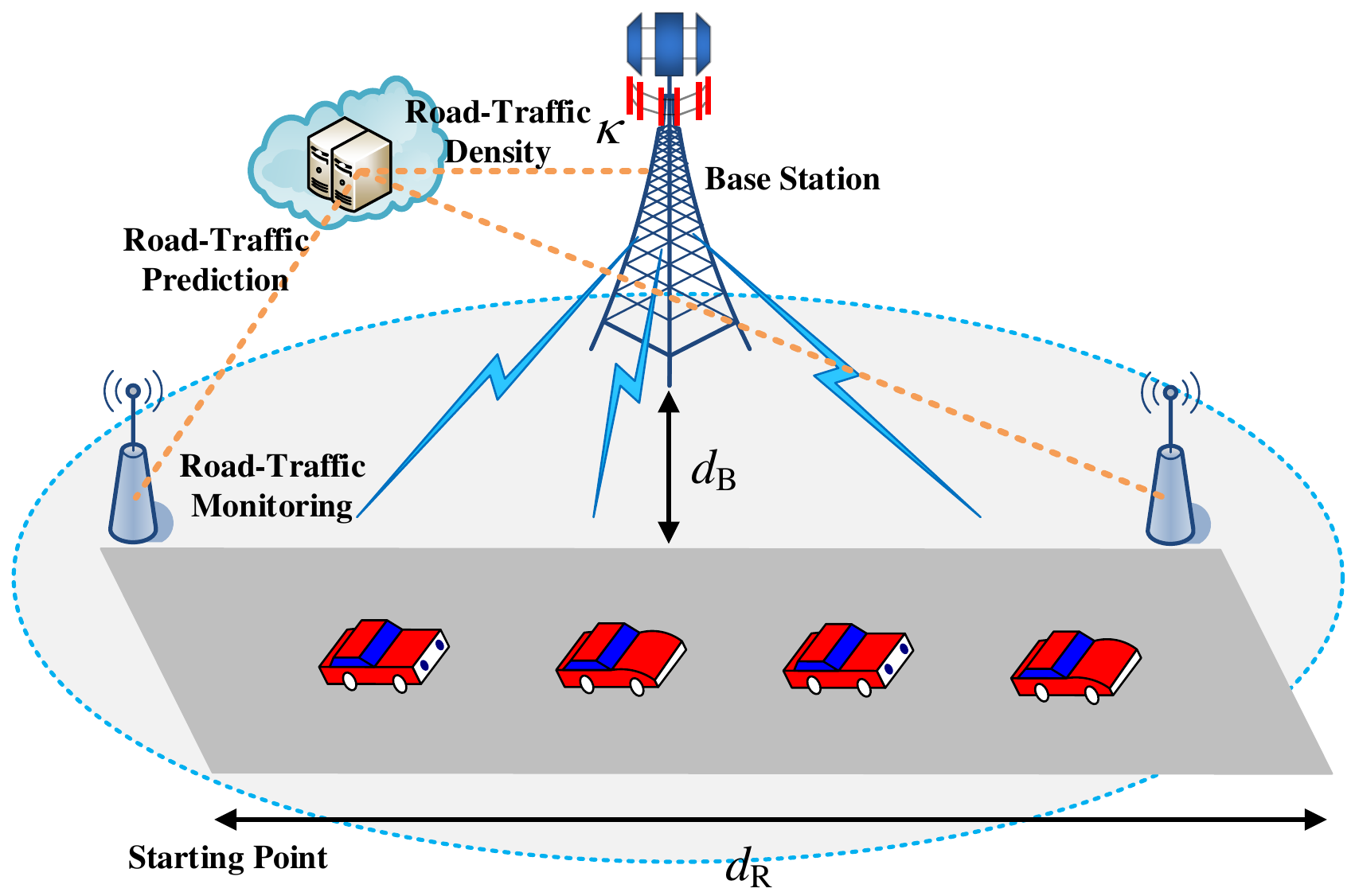}\label{fig_systemcase2}}
		\hfil
		\subfloat[Tradeoff among maximum achievable rate, transmission latency and reliability based on the MF precoder, with $ \kappa = 0.05 $, $ M =300 $, $ B = 200~\text{kHz} $ and the total available power is 10 dBW.]{\includegraphics[scale=0.4]{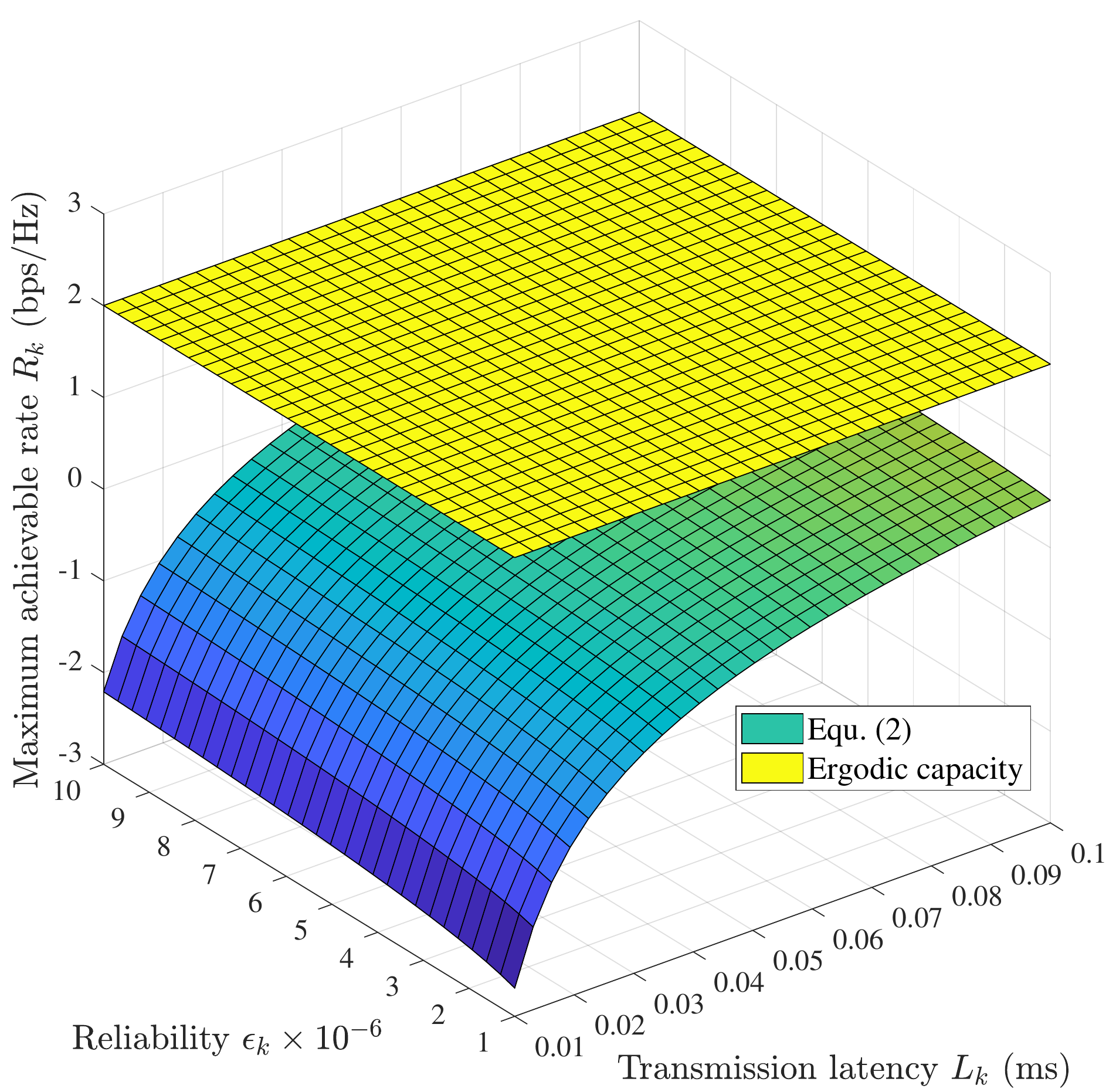}\label{fig_RLR_MF}}
		\hfil
		\subfloat[Maximum transmission latency versus road-traffic density with $ M =300 $, $ R_k(\rho,L,\epsilon) = 100~\text{kbps} $ and $ \epsilon_k = 10^{-6} $ for all vehicular users.]{\includegraphics[scale=0.3]{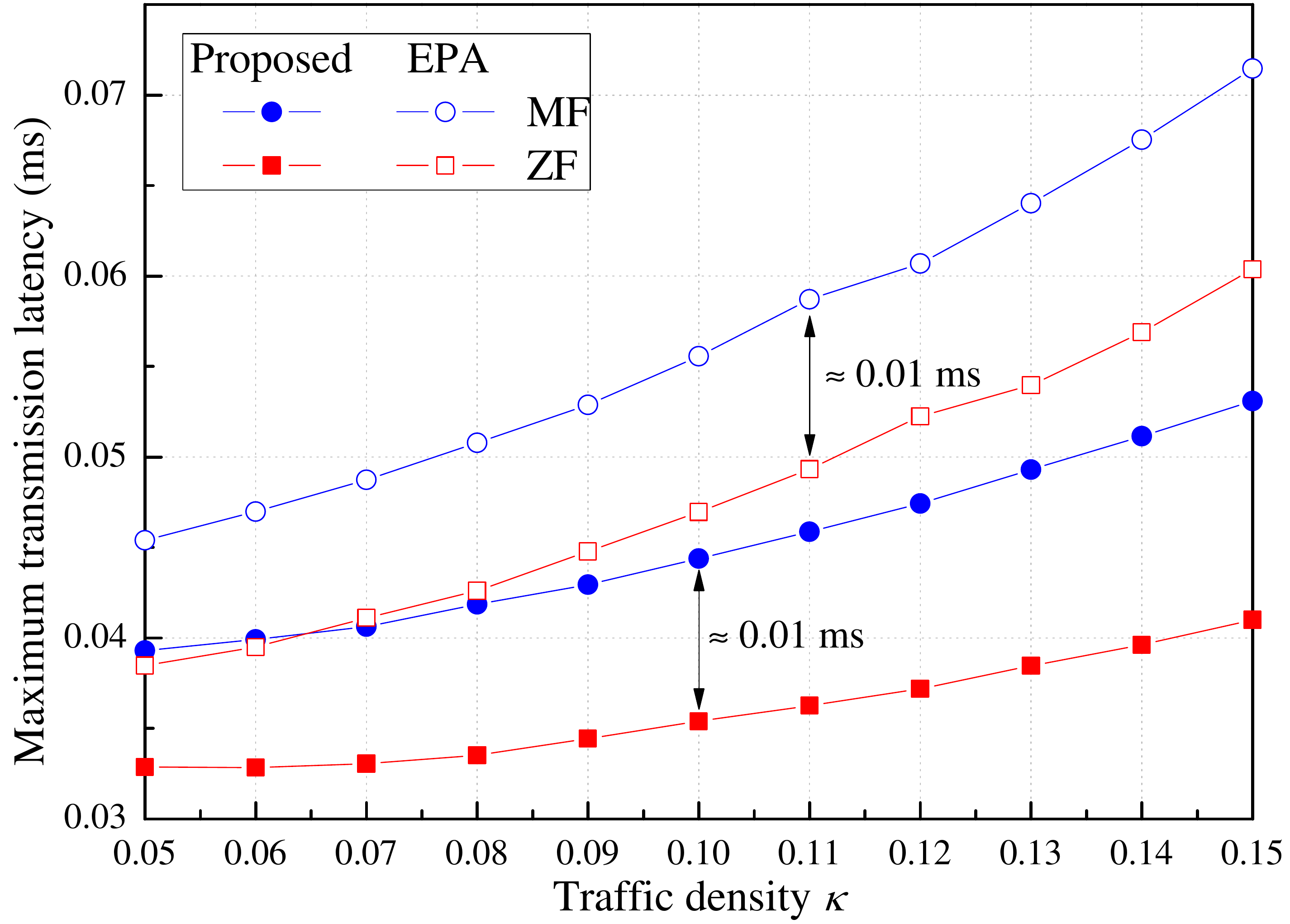}\label{fig_rhovsL}}
		\caption{Case Study~\upperroman{2}---Performance of freeway V2I URLLC system.}
		\label{fig_case2}
	\end{figure}
\fi

According to Case Study~\upperroman{1}, we find that high SNR (or SINR) is conducive to enhance latency and reliability. Massive MIMO relying on hundreds of antennas has been put forward for further improving SNR. Therefore, to exploit the channel hardening phenomenon, a massive MIMO-based scheme is developed for reducing the allocation complexity (only need location information of vehicles) in this case. As shown in Fig.~\subref*{fig_systemcase2}, for the downlink of V2I system aiming for URLLCs, we consider a \textbf{freeway} scenario with a single roadside base station (BS) and a road segment of length $ d_\text{R} $. The BS is $ d_\text{B} $ meters away from the road. The BS employs $ M $ antennas and simultaneously sends information to $ K $ single-antenna aided vehicular users. In order to model the effect of road-traffic on communications, the Underwood \textbf{macroscopic} model is employed and its speed-density function is given by $ v(\kappa) = v_\text{F} \exp\left( -\kappa/\kappa_\text{M} \right) $, where $ v_\text{F} $ is the free-flowing velocity and $ \kappa_\text{M} $ is the maximum density. Then the number of vehicular users can be calculated as $ K = \kappa d_\text{R} $.

We set $ d_\text{B} = 20 $ m, $ d_\text{R} = 200 $ m and $ v_\text{F} = 80 $ km/h. Since the average length of a vehicle is 6.5 m, the maximum road-traffic density is set as $ \kappa_\text{M} = 0.15 $. Perfect CSI is available in this case. Upon using the equal power allocation (EPA) and matched filter (MF) precoder, Fig.~\subref*{fig_RLR_MF} illustrates the tradeoff of a vehicular user among the maximum achievable rate, transmission latency and reliability. As shown in Fig.~\subref*{fig_RLR_MF}, regardless of what the latency and reliability are, the ergodic capacity remains a constant. By contrast, Equ.~\eqref{FBL} indicates that some rate regions are not achievable (negative) because of the ultra-high reliability and ultra-low latency. Moreover, the maximum achievable rate of Equ.~\eqref{FBL} is lower than the ergodic capacity, which confirms that V2I URLLCs are indeed possible, but only at the cost of a reduced rate. Finally, Fig.~\subref*{fig_RLR_MF} shows that the transmission latency can only be optimized within a reasonable range.

In this case, the main objective is to minimize the maximum transmission latency across all vehicular users based on the road-traffic density. Based on the finite blocklength theory, a novel allocation scheme ensuring fairness among all vehicular users is proposed. Fig.~\subref*{fig_rhovsL} illustrates the maximum transmission latency among all vehicular users versus the road-traffic density. As shown in Fig.~\subref*{fig_rhovsL}, when the road-traffic density increases, the maximum transmission latency of both the proposed scheme and EPA scheme increase accordingly. Upon comparing the MF precoder to the zero-forcing (ZF) precoder, we find that the performance of the ZF is better than that of the MF either for the proposed scheme or for the EPA scheme. Furthermore, Fig.~\subref*{fig_rhovsL} indicates that the proposed scheme performs better than the EPA scheme both for the MF and ZF precoders except below $ \kappa = 0.065 $. The performance gap between the MF and ZF schemes is about 0.01 ms. However, the slope of the proposed scheme is significantly lower than that of the EPA scheme, which implies that the proposed scheme is not sensitive to the road-traffic density.


\subsection{Frameworks for MAC Layer}

\subsubsection{Fundamental Tradeoff}

In the system reliability theory, the reliability function is described as a probability that the time to failure (random variable) is greater than a value. Inspired by this thought, the paradigm can be shifted to MAC layer. Then, the reliability is modeled as the probability that the queueing latency does not exceed an expected threshold. Hence, the general expression for the queueing latency $ L $ and reliability $ r(t) $ is given by
\begin{align}
\label{maclatecy}
r(t) &=\Pr \left\{ L\leqslant t \right\} =F_L(t) =\int_{ - \infty }^{t} L f(L) \text{d}L,
\end{align}
where $ F_L(t) $ and $ f(L) $ are the CDF and PDF of the queueing latency, respectively. Moreover, another common metric is the violation probability written as $ \Pr \left\{ L\geqslant t \right\} = 1-r(t) $. According to the general expression, the optimization goal is to let the CDF curve move left as far as possible. Diverse techniques and theories can be used to analyze and optimizec the queueing latency, such as the large deviation theory, stochastic network calculus and Lynapnov optimization. Generally, all of these techniques are based on the queueing theory.


\subsubsection{Large Deviation Theory (Effective Bandwidth or Capacity)}
In general, the optimization objective of latency can be written as $ \Pr \left\{ L \geqslant L_\text{th} \right\} \leqslant \epsilon $, where $ \epsilon $ denotes the reliability requirement. On the basis of the large deviation theory, we have $ \sup \left\lbrace \Pr \left[ L \geqslant L_\text{th} \right] \right\rbrace \approx f\left( R,L_\text{th} \right) \Rightarrow f\left( R,L_\text{th} \right) \leqslant \epsilon $, where $ f\left( R,L_\text{th} \right) $ represents a function about the rate $ R $ determined by the specific scenario. Hence, the objective is equivalently transformed into data rate constraint, i.e., the minimum amount of resources should be allocated to guarantee the latency requirement~\cite{Mei2018}. It is widely exploited that this transformation is only beneficial to the scenario with the large latency regime where the probability of empty buffer is small.

\subsubsection{Stochastic Network Calculus}
The goal of network calculus is to analyze the violation probability $ \Pr \left\{ L\geqslant t \right\} $ (the upper bound of latency). The derivations of stochastic arrival curve (SAC) and stochastic service curve (SSC) play a vital role in the network calculus. Once the SAC and SSC are given, the latency bound can be intuitively obtained by the moment generating function-based and complementary cumulative distribution function-based methods~\cite{Lei2016}. Similarly, let $ \sup \left\lbrace \Pr \left[ L \geqslant L_\text{th} \right] \right\rbrace \leqslant \epsilon $, the tradeoff between latency and reliability is illustrated for further optimizing performance.

\subsubsection{Lynapnov Optimization}
The queue stability is the core metric in the Lyapunov-related theory. Based on the Lyapunov drift theory, the objective of Lyapunov optimization is to maintain the queue stability, and additionally optimize the other performance metrics and satisfy constraints~\cite{Lau2010}. The methodology of Lynapnov optimization is to calculate and optimize the Lyapunov drift with the queue state information (QSI) in the stability region. According to the technique of virtual cost queues, all constraints can be converted into the queue stability problems.


\subsubsection{Case Study~\upperroman{3}---Performance of Urban LTE V2V System~\cite{Mei2018}}

\begin{figure}[!t]
	\centering
	\subfloat[Urban V2V system based on microscopic road-traffic model, where vehicular user and cellular user are abbreviated as VUE and CUE, repectively.]{\includegraphics[scale=0.4]{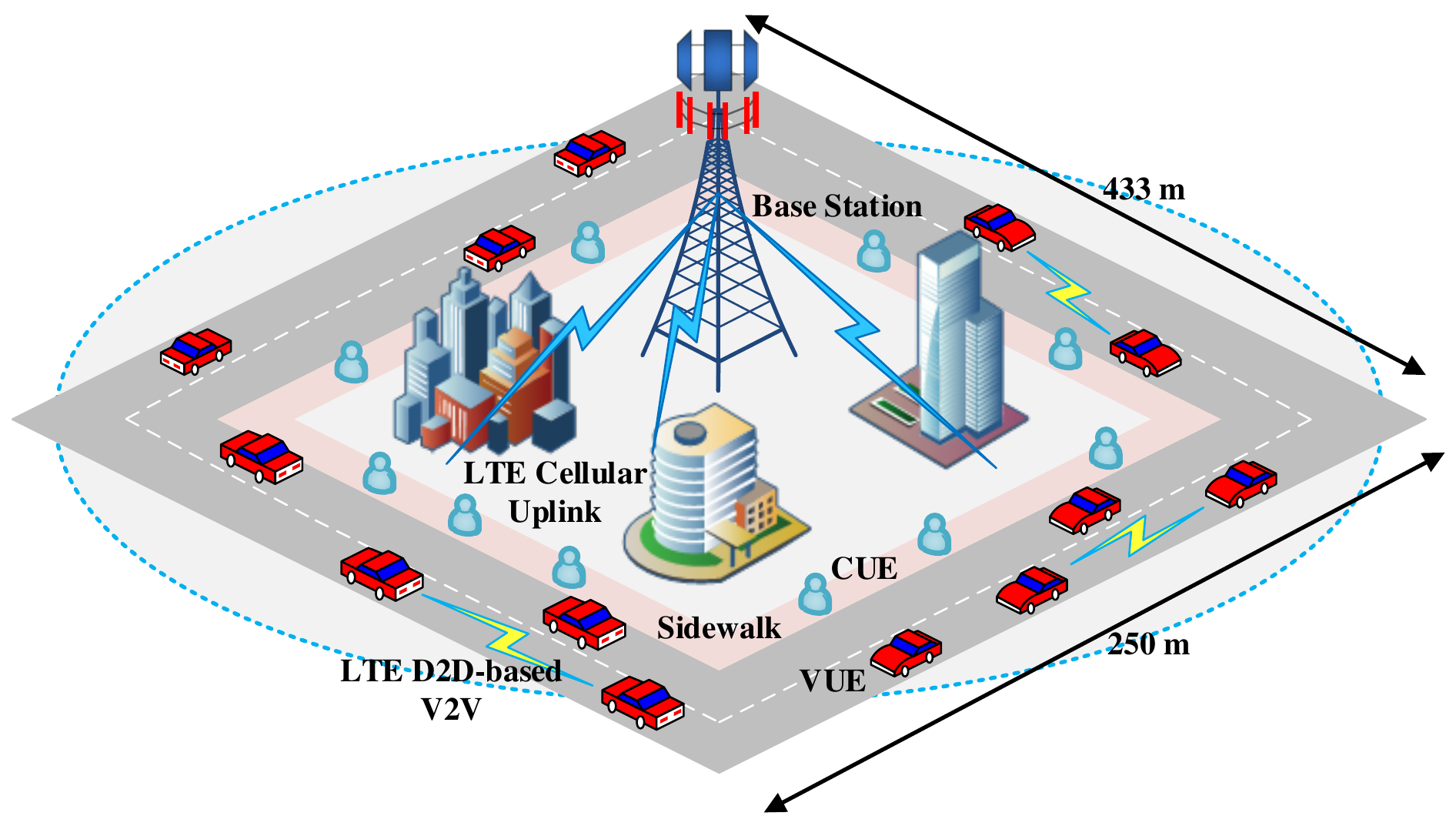}\label{fig_systemcase3}}
	\hfil
	\subfloat[Packet latency PDF of a vehicular user: the $ i $-th bin denotes that the probability of packet latency is larger than $ (i-1) \times 10 $ ms and less than $ i \times 10 $ ms.]{\includegraphics[scale=0.3]{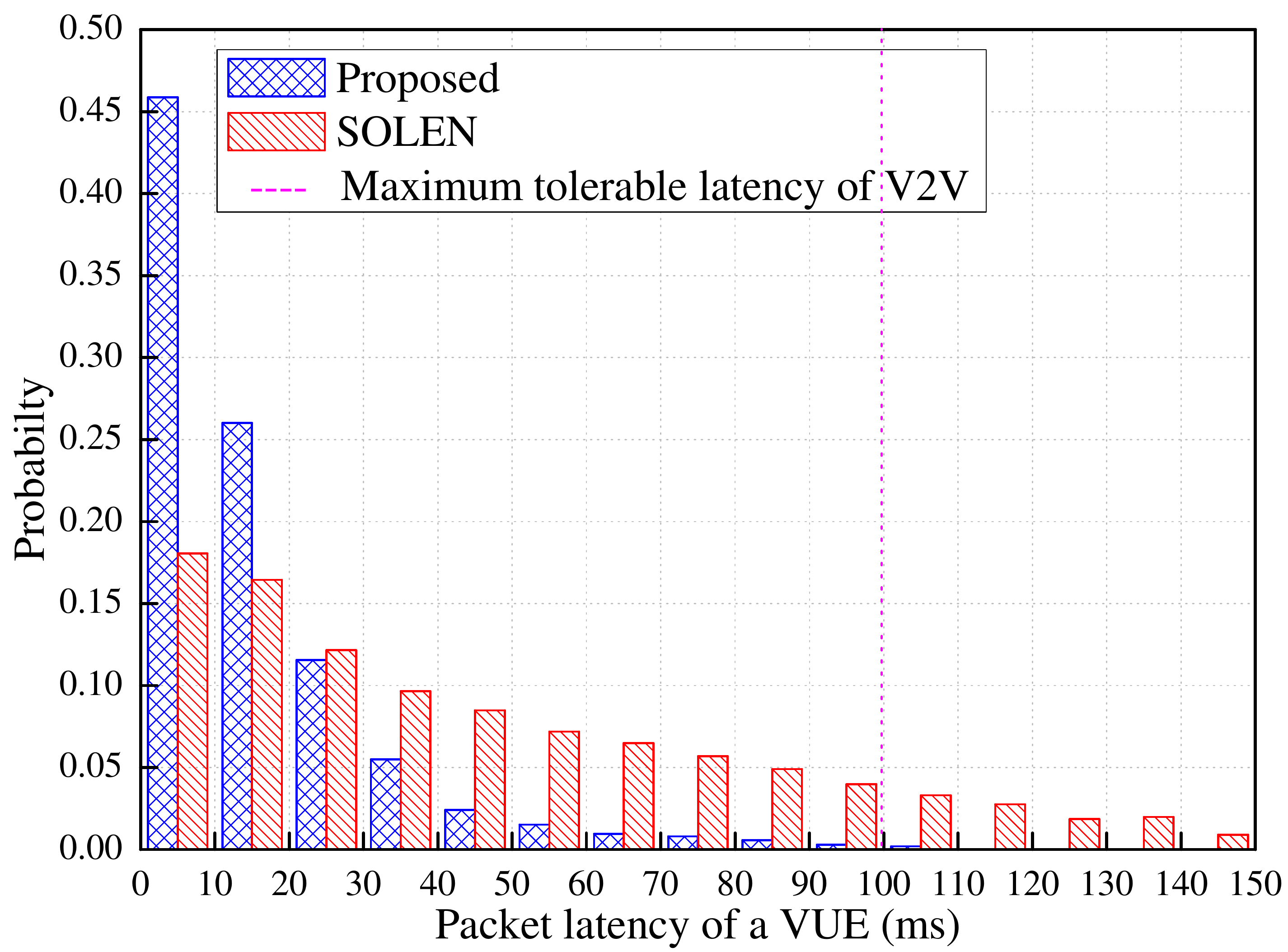}\label{fig_probability}}
	\caption{Case Study~\upperroman{3}---Performance of urban LTE V2V system.}
	\label{fig_case3}
\end{figure}

Nowadays, the cooperation capabilities of vehicles largely depend on timely collecting and sharing of critical information through V2V communications. In order to satisfy the URLLC requirements in MAC layer, the performance of the V2V system is illustrated with the aid of the \textbf{large deviation theory} discussed. As shown in Fig.~\subref*{fig_systemcase3}, based on the Manhattan grid layout, an \textbf{urban} LTE V2V system is considered, where the road grid size is a 433 m $ \times $ 250 m rectangle, and the building size is 413 m $ \times $ 30 m. Furthermore, 3 m is reserved for sidewalk along the building. There are also 2 lanes in each direction, and the lane width is set as 3.5 m. Finally, cellular users are distributed uniformly along the sidewalk, and vehicular users obey the \textbf{microscopic} mobility model described in~\cite{3gpp885}.

In this case, the cellular users share available uplink radio resources with the vehicular users. Our main objective is to maximize the minimum SINR among all cellular users, while meeting the latency and reliability requirements of all vehicular users. We set $ L_\text{th} = 100 $ ms and $ \epsilon = 0.05 $. The performance of the proposed resource management scheme is compared with that of the Separate resOurce bLock and powEr allocatioN (SOLEN) scheme described in~\cite{Mei2018}, which does not consider the effect of packet arrival process on latency. Fig.~\subref*{fig_probability} depicts the PDF of packet latency for the two schemes. In the SOLEN scheme, about 11\% of packet latency exceeds the maximum tolerable packet latency, i.e. 100 ms. This is because the SOLEN scheme only considers the minimum data rate to transmit one packet in its latency transformation method. When the packet arrival rate is low, this static latency transformation method is efficient, however, with the increasing packet arrival rate, this method cannot satisfy the latency requirement of vehicular user due to the backlog of packets in its buffer. Compared to the SOLEN scheme, the packet latency of the proposed scheme is mainly distributed in the range of 0 to 30 ms, and is strictly less than the maximum packet latency. Therefore, the proposed scheme can satisfy the latency requirement of V2V communications, and the techniques and theories discussed before can handle the queueing latency well.

\section{Future Research Directions for URLLC VNETs}
\label{sec:Directions}

In this section, we discuss several open research directions including, but not limited to, the following.

\subsection{Numerology Design}

For the operational cellular systems, the numerology design of OFDM is a very efficient way to reduce latency. In general, the numerology design should be flexible to support various scenarios and requirements. Due to the URLLC requirements, the numerology of URLLCs has the largest subcarrier spacing, which leads to the shortest transmission time interval (TTI). Specifically, 3GPP has agreed that the subcarrier spacing can be chosen based on $ 15 \times 2^n $ kHz, where $ n $ is an integer. However, the choice of $ n $ depends on many factors, such as deployments, mobility and implementation complexity. Therefore, how to design the parameter $ n $ in the actual environments remains as an open problem for V2X communications.

\subsection{Scheduling Schemes}


Each LTE-based V2X transmission is scheduled by base stations or access points with a request-grant procedure, which leads to the extra overhead and latency. Therefore, it is urgent to develop the grant-free multiple access techniques in order to meet the URLLC requirements. In the semi-persistent scheduling (SPS), the resource pools of schedule assignment and data are jointly placed in a certain bandwidth with the frequency division multiplexing manner rather than the time division multiplexing~\cite{Shao2016}. Obviously, the SPS algorithm is more beneficial to decrease latency and improve resource utilization. Because of the characteristics of self-organization in V2X communications, the study of other emerging distributed scheduling algorithms is one of the future research directions. Moreover, how to establish the reasonable and accurate cross-layer optimization frameworks for the transmission and queueing latency is still a challenge in URLLC VNETs.

\subsection{Network Slicing}

Future VNETs require not only the techniques of PHY and MAC layers such as ultra-reliable and low-latency transmission and scheduling, but also the special design of network architecture. The current wireless networks utilize a relatively monolithic framework to carry all kinds of services such as the mobile data from smart phones, high-speed vehicles and embedded machine-to-machine devices. In order to efficiently support vertical industry applications and manage network functionalities, end-to-end network slicing is regarded as a promising solution~\cite{Campolo2017}. In general, the concept of network slicing is derived from network virtualization, where the physical networks are sliced into multiple virtual networks. Each slicing is designed and optimized for specific requirements and service. By taking advantage of the above-mentioned solutions of PHY and MAC layers, three slicing are proposed for meeting different requirements. In detail, Slicing~\#1 is designed with the shortest TTI and grant-free access for driving and road safety as well as cooperative awareness and control. Compared with V2V and V2P, V2I and V2N are mainly centralized types, hence massive MIMO and centralized scheduling schemes should be adopted in Slicing~\#2. In addition, MaaS and social entertainment can be supported by the eMBB slicing (Slicing~\#3). Network slicing can not only maximize the use of the previous infrastructures, but also keep good backward compatibility. Therefore, network slicing provides a feasible evolution roadmap from 4G to 5G, and offers better services to vehicular users. However, it is still an open issue requiring multidisciplinary knowledge and efforts to address.

\section{Conclusions}
\label{sec:Conclusions}

In this article, we investigated the challenges and solutions of conceiving URLLC VNETs dedicated for connected vehicles. In detail, we first introduced various application scenarios based on V2X communications. Then, we discussed the communication requirements and potential challenges. Subsequently, the solutions were presented from the aspects of PHY and MAC layers. Particularly, the optimization frameworks for latency and reliability, and the case studies of the freeway V2I and urban V2V systems were investigated. Finally, several future research directions for URLLC VNETs were discussed. We hope this article can inspire more ground-breaking research efforts along this emerging branch in the future.

%
\bibliographystyle{IEEEtran}
\bibliography{IEEEabrv,Ref}

\begin{thebibliography}{10}
\providecommand{\url}[1]{#1}
\csname url@samestyle\endcsname
\providecommand{\newblock}{\relax}
\providecommand{\bibinfo}[2]{#2}
\providecommand{\BIBentrySTDinterwordspacing}{\spaceskip=0pt\relax}
\providecommand{\BIBentryALTinterwordstretchfactor}{4}
\providecommand{\BIBentryALTinterwordspacing}{\spaceskip=\fontdimen2\font plus
\BIBentryALTinterwordstretchfactor\fontdimen3\font minus
  \fontdimen4\font\relax}
\providecommand{\BIBforeignlanguage}[2]{{%
\expandafter\ifx\csname l@#1\endcsname\relax
\typeout{** WARNING: IEEEtran.bst: No hyphenation pattern has been}%
\typeout{** loaded for the language `#1'. Using the pattern for}%
\typeout{** the default language instead.}%
\else
\language=\csname l@#1\endcsname
\fi
#2}}
\providecommand{\BIBdecl}{\relax}
\BIBdecl

\bibitem{Siegel2018}
J.~E. {Siegel}, D.~C. {Erb}, and S.~E. {Sarma}, ``A survey of the connected
  vehicle landscape--{Architectures}, enabling technologies, applications, and
  development areas,'' \emph{{IEEE} Trans. Intell. Transp. Syst.}, vol.~19,
  no.~8, pp. 2391--2406, Aug. 2018.

\bibitem{adv2015}
K.~Zheng, Q.~Zheng, H.~Yang \emph{et~al.}, ``Reliable and efficient autonomous
  driving: {T}he need for heterogeneous vehicular networks,'' \emph{{IEEE}
  Commun. Mag.}, vol.~53, no.~12, pp. 72--79, Dec. 2015.

\bibitem{3gpp885}
``Study on {LTE}-based {V2X} services,'' 3GPP, Tech. Rep. TR 36.885 V14.0.0,
  Jun. 2016.

\bibitem{itu2015}
\emph{{IMT Vision - Framework and overall objectives of the future development
  of IMT for 2020 and beyond}}, ITU-R Std. M.2083-0, Sep. 2015.

\bibitem{Luan2015}
T.~H. Luan, R.~Lu, X.~S. Shen \emph{et~al.}, ``Social on the road: {E}nabling
  secure and efficient social networking on highways,'' \emph{{IEEE} Wireless
  Commun.}, vol.~22, no.~1, pp. 44--51, Feb. 2015.

\bibitem{3gpp913}
``Study on scenarios and requirements for next generation access
  technologies,'' 3GPP, Tech. Rep. TR 38.913 V15.0.0, Jun. 2018.

\bibitem{Long2018}
H.~{Long}, W.~{Xiang}, and Y.~{Sun}, ``Low-latency and high-reliability
  performance analysis of relay systems,'' \emph{IET Commun.}, vol.~12, no.~5,
  pp. 627--633, Mar. 2018.

\bibitem{Polyanskiy2010}
Y.~Polyanskiy, H.~V. Poor, and S.~Verdu, ``Channel coding rate in the finite
  blocklength regime,'' \emph{{IEEE} Trans. Inf. Theory}, vol.~56, no.~5, pp.
  2307--2359, May 2010.

\bibitem{She2017}
C.~She, C.~Yang, and T.~Q.~S. Quek, ``Radio resource management for
  ultra-reliable and low-latency communications,'' \emph{{IEEE} Commun. Mag.},
  vol.~55, no.~6, pp. 72--78, Jun. 2017.

\bibitem{Yang2019}
H.~{Yang}, K.~{Zheng}, L.~{Zhao} \emph{et~al.}, ``Twin-timescale radio resource
  management for ultra-reliable and low-latency vehicular networks,''
  \emph{{\tt arXiv:1903.01604v1 [cs.IT]}}, pp. 1--30, Mar. 2019.

\bibitem{Mei2018}
J.~Mei, K.~Zheng, L.~Zhao \emph{et~al.}, ``A latency and reliability guaranteed
  resource allocation scheme for {LTE V2V} communication systems,''
  \emph{{IEEE} Trans. Wireless Commun.}, vol.~17, no.~6, pp. 3850--3860, Jun.
  2018.

\bibitem{Lei2016}
L.~{Lei}, J.~{Lu}, Y.~{Jiang} \emph{et~al.}, ``Stochastic delay analysis for
  train control services in next-generation high-speed railway communications
  system,'' \emph{{IEEE} Trans. Intell. Transp. Syst.}, vol.~17, no.~1, pp.
  48--64, Jan. 2016.

\bibitem{Lau2010}
V.~K.~N. Lau and Y.~Cui, ``Delay-optimal power and subcarrier allocation for
  {OFDMA} systems via stochastic approximation,'' \emph{{IEEE} Trans. Wireless
  Commun.}, vol.~9, no.~1, pp. 227--233, Jan. 2010.

\bibitem{Shao2016}
S.~Sun, J.~Hu, Y.~Peng \emph{et~al.}, ``Support for vehicle-to-everything
  services based on {LTE},'' \emph{{IEEE} Wireless Commun.}, vol.~23, no.~3,
  pp. 4--8, Jun. 2016.

\bibitem{Campolo2017}
C.~{Campolo}, A.~{Molinaro}, A.~{Iera} \emph{et~al.}, ``{5G} network slicing
  for vehicle-to-everything services,'' \emph{{IEEE} Wireless Commun.},
  vol.~24, no.~6, pp. 38--45, Dec. 2017.

\end{thebibliography}
%
%
\end{document}